\documentclass[aps,prd,11pt,superscriptaddress,amssymb,amsmath,nofootinbib]{revtex4-1}

\usepackage{hhline}
\usepackage{color}
\usepackage{epsfig}
\usepackage{graphicx}
\usepackage{float}
\usepackage{epstopdf}
\usepackage{hyperref}
\usepackage[usenames,dvipsnames]{xcolor}
\usepackage{multirow}
\usepackage[normalem]{ulem}
\usepackage[d]{esvect}

\begin{document}

 \title{Matter effects on flavor transitions of high-energy astrophysical neutrinos based on typical propagation schemes }
 \author{Ding-Hui Xu}\email{3036602895@qq.com}
 \author{Shu-Jun Rong}\email{rongshj@glut.edu.cn}
 \affiliation{College of Physics and Electronic Information Engineering, Guilin University of Technology, Guilin, Guangxi 541004, China}

 \begin{abstract}
Precise measurements of the flavor ratio of high energy astronomical neutrinos (HANs) would be promising in the following decades.
Then matter effects and new physics effects on the flavor transition of HANs could be tested. In this paper, we examine matter effects on the flavor composition of HANs.
The effects are dependent on propagation schemes and sources of neutrinos. We consider propagations in environments with adiabatically varying and constant electron density.
In the adiabatic case, the matter influence on the flavor composition of HANs at Earth is noticeable at the electron density 10$^{10}$ cm$^{-3}$ in the cases of muon damping and neutron decaying sources. In contrast, in the constant density case, the matter effect can be neglected even at the density 10$^{18}$ cm$^{-3}$. Thus, observations from next-generation neutrino telescopes may set stringent constraints on the adiabatic propagation scheme.

\end{abstract}

 \maketitle

 \section{Introduction}

In the recent decade, many high-energy astrophysical neutrinos (HANs) events in the TeV - PeV energy range have been observed by the IceCube observatory\cite{1,2,3,4,5}.
The energy spectrum, flavor ratio, and sky distribution of HANs are analysed\cite{6,7,8,9,10}.
With the help of the synergetic observation of electromagnetic wave messengers, the promising sources of HANs are indicated\cite{11,12,13,14}.
Although considerable progresses have been made in neutrino astronomy, there are several important problems to be solved: The uncertainty of flavor ratio measured at Earth is large\cite{15,16}; No source of HANs is identified above the $5\sigma$ discovery threshold; How to explain the multi-messenger emission from the blazar TXS0506+056 is challenging\cite{17,18,19}; The production mechanism of HANs is unclear. Among these problems, the first one would be overcome with the operations of the next-generation neutrino telescopes\cite{20,21,22,23,24} in the following decades.
Furthermore, since the energy spectrum and the flavor ratio of HANS at the source are dependent on the production mechanism, the progress from the first one would also provide clues for the last one.

As is known, the flavor ratio at the source is inferred from the flavor ratio at Earth and the flavor transition probability of HANs. Under the assumption that
the leptonic mixing matrix at the source and the matrix in the propagation path are the same as the one in vacuum,
the so-called standard flavor transition matrix (FTM) is obtained after the decoherence of neutrinos, i.e.,
\begin{equation}
\label{eq:1}
\overline{P}_{\alpha\beta}^{s}=\sum_{i}|U_{\alpha i}|^{2}|U_{\beta i}|^{2},
\end{equation}
where $U$ is the leptonic mixing matrix in vacuum with $\alpha, \beta = e, \mu, \tau$, $i = 1, 2, 3$\cite{25}.
At present, the constraint on the FTM from the measured flavor ratio is loose. So this minimal scenario works well. However, precise measurements of the flavor ratio would be available in the future.
Then a scenario including matter effects or new physics effects could be tested. Various FTMs based on new physics have been proposed, see \cite{26,27,28,29,30,31,32,33,34,35,36,37,38,39,40,41,42,43,44,45,46,47} for example.
In this paper, we are concerned with  impacts of  matter on the flavor transition of HANs.

  In general, the matter effects on neutrinos arise from coherent forward elastic and incoherent scatterings of HANs
with  medium particles. The former affects the flavor conversion through the neutrino mixing matrix in matter. The latter changes not only the flavor composition  but also the energy spectrum of HANs.
It plays an important role in the recombination  process of HANs in dense medium\cite{48}. We focus on the matter effects from the elastic scattering in following sections.
Considering the uncertainties in energy of HANs, the size of the medium, and the profile of the electron density, the FTMs are dependent on the propagation schemes in matter.
Even for the same medium environment, different interpretations on decoherence of neutrinos in propagations lead to different FTMs\cite{49,50}.
In this paper, we consider two typical schemes, namely propagation in the medium with an adiabatically varying electron density  and  propagation in  medium with a constant density. We assume that the propagation distance
is much larger than the oscillation length both in matter and in vacuum. The neutrino oscillation probabilities are averaged to obtain FTMs. Based on the ansatz, influences of electron density on the flavor ratio of HANs at Earth could be compared in the schemes.

The paper is organised as follows. In Sec. \uppercase\expandafter{\romannumeral2}, we describe the mixing matrix and oscillation probability of neutrino in matter.
The FTMs are obtained from the propagation schemes proposed in Refs.\cite{51,52} and a modified condition.
In Sec. \uppercase\expandafter{\romannumeral3}, we perform a binned maximum likelihood analysis to constrain the electron density in the neutrino propagation path.
The impacts of matter on the flavor ratio of HANs at Earth are examined with three typical sources at low and high electron densities. Finally, we conclude.

\section{FTMs based on typical propagation schemes}

\subsection{The mixing matrix and oscillation probabilities of neutrinos in matter}
In the standard three flavor framework, the effective Hamiltonian $H^{m}$ in the flavor basis responsible for the propagation of neutrinos in matter is expressed as\cite{53,54}
\begin{equation}
\label{eq:2}
H^{m}=\frac{1}{2E}U
\left(
    \begin{array}{ccc}
    0&0&0\\
    0&\Delta m^{2}_{21}&0\\
    0&0&\Delta m^{2}_{31}\\
    \end{array}
  \right)U^{\dagger}\pm\sqrt{2}G_{F}
\left(
    \begin{array}{ccc}
    N_{e}&0&0\\
    0&0&0\\
    0&0&0\\
    \end{array}
  \right)=U^{m}
  \left(
    \begin{array}{ccc}
    E^{m}_{1}&0&0\\
    0&E^{m}_{2}&0\\
    0&0&E^{m}_{3}\\
    \end{array}
  \right)U^{m^{\dagger}}.
\end{equation}
Here $E$ is the neutrino energy, $G_{F}$ is the Fermi constant, $\Delta m^{2}_{21}$ and $\Delta m^{2}_{31}$ are the neutrino mass-squared differences, and $N_{e}$ is the electron density. For antineutrinos, all terms should be replaced by their complex conjugates and the sign of $N_{e}$ is negative.
The leptonic mixing matrix in matter, $U^{m}$, can be obtained by the diagonalization of $H^{m}$. The analytical expression of $U^{m}$ is given as the Ref.\cite{55}.
In the medium with a constant electron density, the neutrino oscillation probability  is expressed as follow
\begin{equation}
\label{eq:3}
P^{m}_{\alpha \beta}=\delta_{\alpha \beta}-4\sum_{j>i}Re(U^{m}_{\beta j}U^{m^{\ast}}_{\beta i}U^{m^{\ast}}_{\alpha j}U^{m}_{\alpha i})sin^{2}(\frac{L\Delta E^{m}_{ji}}{2})+2\sum_{j>i}Im(U^{m}_{\beta j}U^{m^{\ast}}_{\beta i}U^{m^{\ast}}_{\alpha j}U^{m}_{\alpha i})sin(L\Delta E^{m}_{ji}),
\end{equation}
where $\Delta E^{m}_{ji}=~E^{m}_{j}-E^{m}_{i}$ is the difference between the eigenvalues of $H^{m}$.

\subsection{FTMs based on two typical propagation schemes}
By now, the production mechanism of HANs is not determined. The matter parameters  of the environment around the HAN source are uncertain. Accordingly, the propagation scheme and the FTM of HANs
are not definite. Here we consider two typical schemes which may model some realistic situations undergone by HANs.

First we suppose that neutrinos propagate in the medium  where the transitions between the Hamiltonian eigenstates in matter are negligible.
Moreover, the propagation distance considered is much larger then the oscillation length both in matter and in vacuum. Similar to the case of solar neutrinos, HANs arrive at Earth as an incoherent sum of the three mass eigenstates. In this case, we can obtain an adiabatic FTM of the form\cite{52}
\begin{equation}
\label{eq:4}
\overline{P}^{\uppercase\expandafter{\romannumeral1}}_{\alpha\beta}=\sum_{i}|U_{\alpha i}^{m}|^{2}|U_{\beta i}|^{2},
\end{equation}
where $U^{m}$ is the leptonic mixing matrix in matter.

As a comparison, in this paper we also consider another scheme where the electron density is constant in matter.
The effective decoherence of neutrinos is  obtained from the integral of the oscillation probability over the whole propagation path, i.e.\cite{51},
\begin{equation}
\label{eq:5}
\overline{P}^{\uppercase\expandafter{\romannumeral2}}_{\alpha\beta}=\int_{0}^{L_{vac}}\int_{0}^{L}P_{\alpha\beta}(l_{vac},l_{m})dl_{vac}dl_{m}L_{vac}^{-1}L^{-1},
\end{equation}
with
\begin{equation}
\label{eq:6}
\begin{aligned}
P_{\alpha\beta}(l_{vac},l_{m})&=|\langle\nu_{\beta}|Ue^{-iH_{vac}l_{vac}}U^{+}U^{m}e^{-iH_{m}l_{m}}U^{m +}|\nu_{\alpha} \rangle|^{2}\\
&=|\sum_{\gamma}\sum_{i}\sum_{j}U_{\beta i}e^{-iE_{i}l_{vac}}U_{\gamma i}^{*}U_{\gamma j}^{m}e^{-iE_{j}^{m}l_{m}}U_{\alpha j}^{m*}|^{2}\\
&=\sum_{\gamma}\sum_{\gamma'}\sum_{i}\sum_{i'}\sum_{j}\sum_{j'}U_{\beta i}U_{\beta i'}^{*}e^{-i(E_{i}-E_{i'})l_{vac}}U_{\gamma i}^{*}U_{\gamma' i'}U_{\gamma j}^{m}U_{\gamma' j'}^{m*}e^{-i(E_{j}^{m}-E_{j'}^{m})l_{m}}
U_{\alpha j}^{m*}U_{\alpha j'}^{m},
\end{aligned}
\end{equation}
where $L$ ($L_{vac}$) is the propagation distance in matter (vacuum), $H_{m}$ ($H_{vac}$) is the diagonal neutrino Hamiltonian matrix in matter (vacuum).
When the distance significantly exceeds the oscillation length, i.e., $L\gg L^{osc}$, $L_{vac}\gg L^{osc}_{vac}$, the oscillation terms disappear and the averaged oscillation probability can
be written as follow
\begin{equation}
\label{eq:7}
\overline{P}^{\uppercase\expandafter{\romannumeral2}}_{\alpha\beta}=\sum_{i}\sum_{j}|U^{m}_{\alpha j}|^{2}P_{ji}|U_{\beta i}|^{2},
\end{equation}
with
\begin{equation}
\label{eq:8}
P_{ji}=|\sum_{\gamma }U^{*}_{\gamma i}U_{\gamma j}^{m}|^{2}.
\end{equation}
Note that, after the averaging, the transition probability $\overline{P}^{\uppercase\expandafter{\romannumeral2}}_{\alpha\beta}$ in Eq.\ref{eq:7} is different from
the one given in Ref.\cite{51} where the condition  $L\gg L^{osc}$ is not respected.

\section{Matter effects on the flavor ratio}
\subsection{ Likelihood analysis on the electron density}
In order to examine the matter effects on the flavor ratio of HANs, we first perform a simplified analysis on the electron density in the environment around the HAN source, employing the high-energy starting events (HESEs) detected by the IceCube observatory.
With respect to the data sample, 4318 days of HESEs are employed, with 164 updated events\cite{IceCube:2023sov}.
Since there may be background events in the sample at low energies \cite{4,Halzen:2016pwl,Halzen:2016thi} and the track event number is negligible at $E_{\nu}>$210TeV, we set the energy range of neutrinos to 55$\div$210 TeV with a total of 81 data samples, see Tab.\ref{Tab:1}.
\begin{table}
	\caption{\label{Tab:1} The shower and track event number in the chosen energy range.}
	\centering
	\begin{tabular}{cccccccc}
		\hline
		Energy range / Tev~&~55$\div$69~&~69$\div$83~&~83$\div$100~&~100$\div$120~&~120$\div$144~&~144$\div$1174~&~174$\div$209\\
		\hline
		Shower event number~&7~&~7~&~10~&~8~&~11~&~7~&~3\\
		Track event number~~&5~&~10~&~2~&~6~&~2~&~2~&~1\\
		\hline
	\end{tabular}
\end{table}

We use a binned maximum likelihood method to constrain the electron density in the medium.
Considering the dominant contribution of the charged current to the HESEs\cite{35}, the expected shower and track event number at each energy bin k are given respectively
\begin{equation}
\label{eq:9}
	N^{sh}_{k}(N_{e})=4\pi T(\int_{k}\Phi_{\nu_{e}+\overline{\nu}_{e}}(E,N_{e})A_{e,k}(E)dE+\int_{k}\Phi_{\nu_{\tau}+\overline{\nu}_{\tau}}(E,N_{e})A_{\tau,k}(E)dE),
\end{equation}
\begin{equation}
\label{eq:10}
	N^{tr}_{k}(N_{e})=4\pi T(\int_{k}\Phi_{\nu_{\mu}+\overline{\nu}_{\mu}}(E,N_{e})A_{\mu,k}(E)dE),
\end{equation}
where T=4318 days, $A_{\alpha,~k}(E)$ denotes the effective area of the IceCube Collaboration for the $\alpha$-flavor neutrino at the k energy interval\cite{IceCube:2013low}, and $\Phi_{\nu_{\alpha}+\overline{\nu}_{\alpha}}(E,N_{e})=\Phi_{\nu+\overline{\nu}}(E)(\sum_{\beta}\overline{P}_{\alpha\beta}(E,N_{e})\phi^{s}_{\beta})$ with $\phi^{s}_{\beta}$ representing the flavor ratio at the source of HANs.
$\Phi_{\nu+\overline{\nu}}(E)$ is the entire diffuse flow, and its specific expression is as follow
\begin{equation}\label{eq:11}
	\Phi_{\nu+\overline{\nu}}(E)=\phi\times(\frac{E}{100~\rm TeV})^{-\gamma}\cdot10^{-18}~\rm GeV^{-1} cm^{-2} s^{-1} sr^{-1}.
\end{equation}
The best fit values and 68\% confidence level of the parameters in the all-sky model are listed in Tab.\ref{Tab:2}\cite{IceCube:2020wum}.
\begin{table}
	\caption{\label{Tab:2} The best fit values and 68\% confidence level of energy-spectrum parameters in the all-sky model\cite{IceCube:2020wum}.}
	\centering
	\begin{tabular}{ccc}
		\hline
		~~~~~~parameters~~~~~~&~~~~~~best fit value~~~~~~&~~~~~~68\% confidence level(C.L.)~~~~~~\\
		\hline
		~~~~~~$\phi$~~~~~~&~~~~~~6.37~~~~~~&~~~~~~~~4.75~-~7.83~~~~~~\\
		~~~~~~$\gamma$~~~~~~&~~~~~~2.87~~~~~~&~~~~~~~~2.68~-~3.08~~~~~~\\
		\hline
	\end{tabular}
\end{table}
The number of shower(track) events in the energy interval k is denoted by $\overline{N^{sh}_{k}}$($\overline{N^{tr}_{k}}$). We assume that the expected event numbers $N^{sh/tr}_{k}$ in different energy intervals k follows the Poisson distribution\cite{56}, namely
\begin{equation}\label{eq:12}
	P[\overline{N^{x}_{k}}|N^{x}_{k}(N_{e})]=\frac{N^{x}_{k}(N_{e})^{\overline{N^{x}_{k}}}}{\overline{N^{x}_{k}}!}\exp\{-N^{x}_{k}(N_{e})\},
\end{equation}
with x=~\{sh,~ tr\}.

The likelihood function is
\begin{equation}
\label{eq:13}
	L(N_{e})=\prod_{k}P[~\overline{N^{sh}_{k}}|N^{sh}_{k}(N_{e})]\times P[~\overline{N^{tr}_{k}}|N^{tr}_{k}(N_{e})].
\end{equation}
Taking the logarithm of both sides of the above equation, we define the function $\chi^{2}$ as follow
\begin{equation}\label{eq:14}
	\chi^{2}\equiv-2(\ln[L(N_{e})]-\ln[L(N_{e}=0)]).
\end{equation}
The function $\chi^{2}$ is dependent not only on the electron density but also on leptonic mixing parameters and the energy-spectrum parameters.
However, we find that $\chi^{2}$ varies slowly with the variable $N_{e}$ irrespective of the values of the mixing and energy-spectrum parameters.
Thus, the best fit data of the oscillation parameters from the global fit NuFIT6.0 \cite{Esteban:2024eli} with  the normal mass-ordering (NO) and the best fit value of the spectrum index in Tab.\ref{Tab:2} are taken  to show the influence of $N_{e}$ in our simplified analysis. Furthermore, the influence of the normalization parameter $\phi$ on $\chi^{2}$ is averaged within the 68\% confidence range, using a Monte Carlo integration method. Based on the set-up of the nuisance parameters, the dependence of $\chi^{2}$ on $N_{e}$ is shown in Fig.\ref{fig:1}.
\begin{figure}\label{fig:1}
	\centering
	\includegraphics[width=0.32\textwidth]{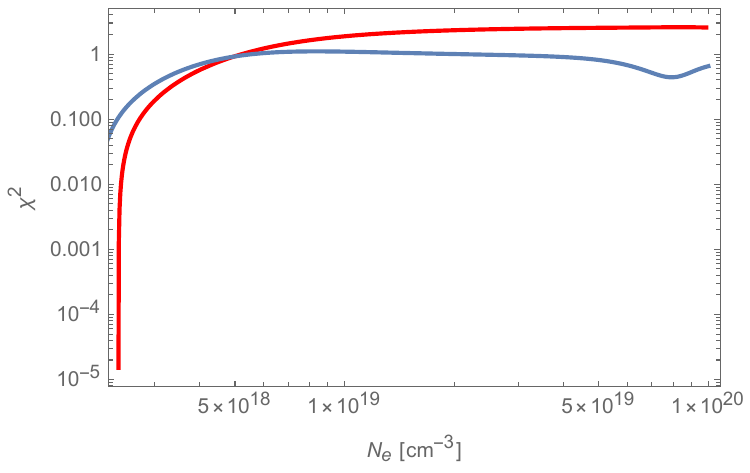}
	\hfill
	\includegraphics[width=0.32\textwidth]{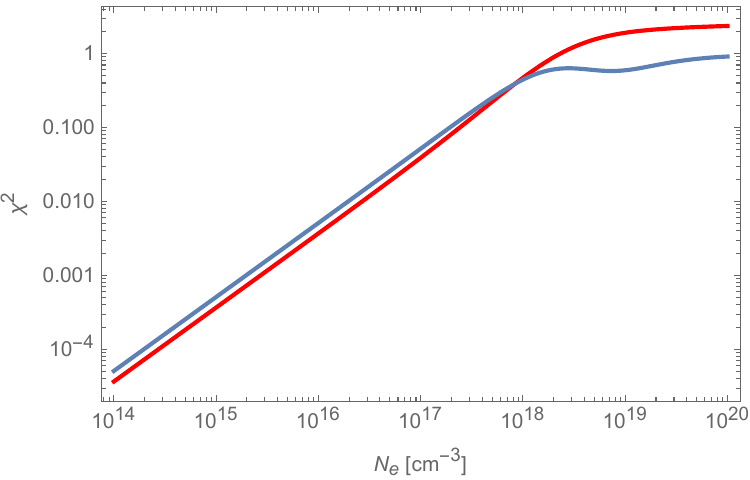}
	\hfill
	\includegraphics[width=0.32\textwidth]{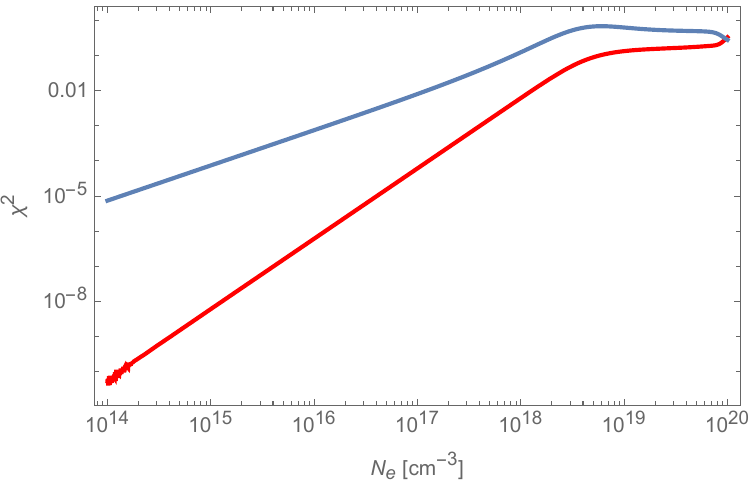}
	\hfill
	\caption{\label{fig:1} The dependence of the function $\chi^{2}$ on  $N_{e}$. The left panel: case of $\mu^{\pm}$ damping source with $\phi^{S}=(0,1,0)$;
The middle panel: case of neutron decaying source with  $\phi^{S}=(1,0,0)$; The right panel: case of $\pi^{\pm}$ decaying source with $\phi^{S}=(1/3,2/3,0)$.
The red lines: results from the Scheme \uppercase\expandafter{\romannumeral1}; The blue lines: results from the Scheme \uppercase\expandafter{\romannumeral2}.}
	\label{fig:1}
\end{figure}

As is displayed in the plots, in the range $N_{e}<10$$^{17}$cm$^{-3}$, we have $\chi^{2}(N_{e})\sim\chi^{2}(N_{e}=0)$. $\chi^{2}\geq1$ is realized in the left/middle panel when $N_{e}$ is of the order of 10$^{18}$cm$^{-3}$. In the case of $\pi^{\pm}$ decaying source, $\chi^{2}\sim0.1$, when $N_{e}$ is as large as 10$^{18}$cm$^{-3}$. Thus, although the samples from the 11.8 years observations are taken, these curves of $\chi^{2}$ show that the magnitude of $N_{e}$ is weakly constrained by the number of track and shower events. In fact, the observation holds despite of the uncertainties of the nuisance parameters in their 1$\sigma$ allowed ranges.

\subsection{Matter effects on the flavor ratio at Earth}
In the previous analysis, we find that the constraint from the present HESE samples on the electron density in the considered schemes is loose.
Thus, a large range of $N_{e}$ can be proposed to check the influence on the flavor ratio of HANs. As illustrative examples, at the level $\chi^{2}<1.5$ in each scheme,
we considerer a low and a high density range, namely [10$^{10}$, ~2$\times10^{10}$]cm$^{-3}$ and [5$\times10^{18}$, ~6$\times10^{18}$]cm$^{-3}$.
Based on the $\mu^{\pm}$ damping, neutron decaying, and $\pi^{\pm}$ decaying source, the ternary plots of the flavor ratio of HANs at Earth are shown in Fig.\ref{fig:2}-Fig.\ref{fig:4}.
\begin{figure}\label{fig:2}
	\centering
	\includegraphics[width=0.45\textwidth]{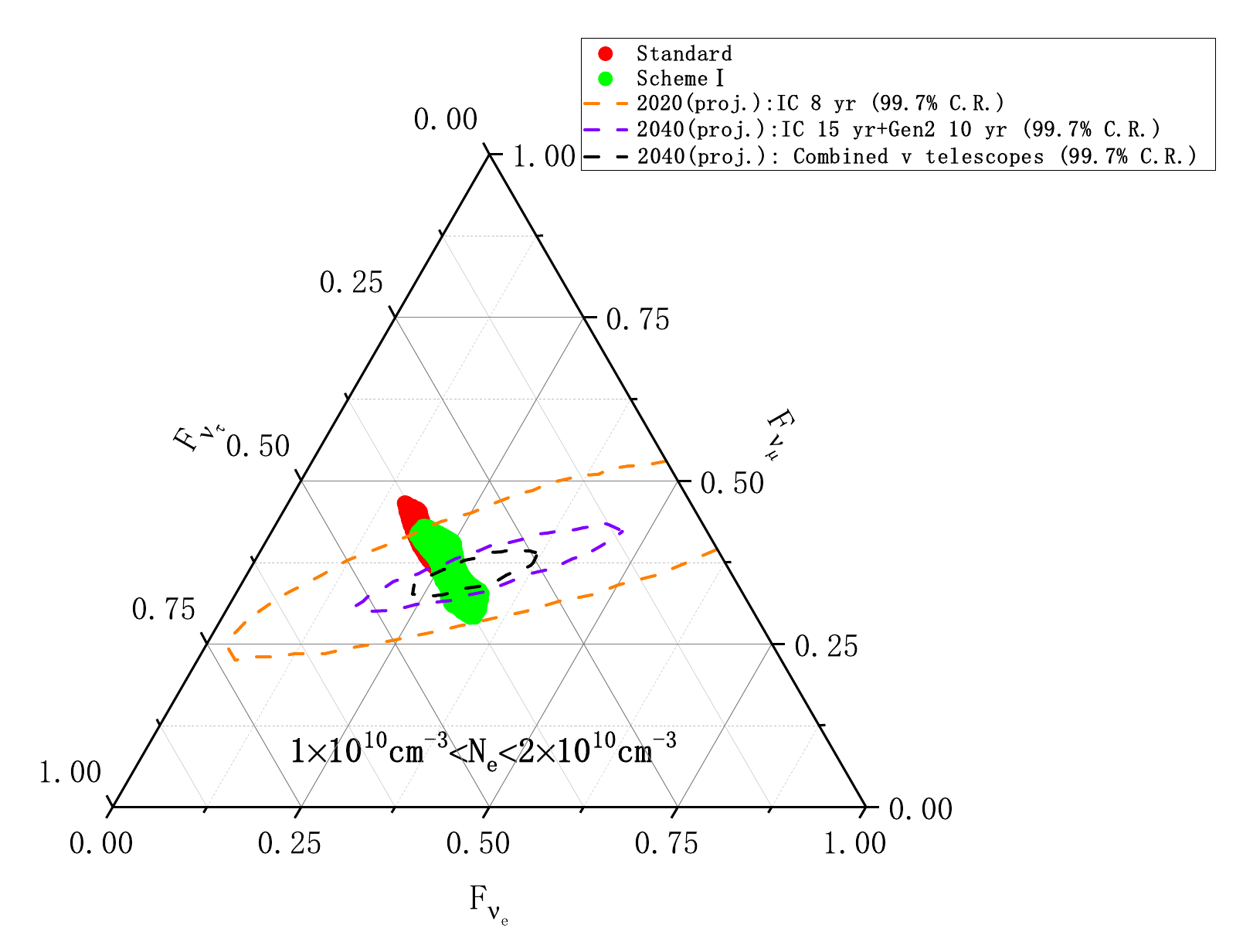}
	\includegraphics[width=0.45\textwidth]{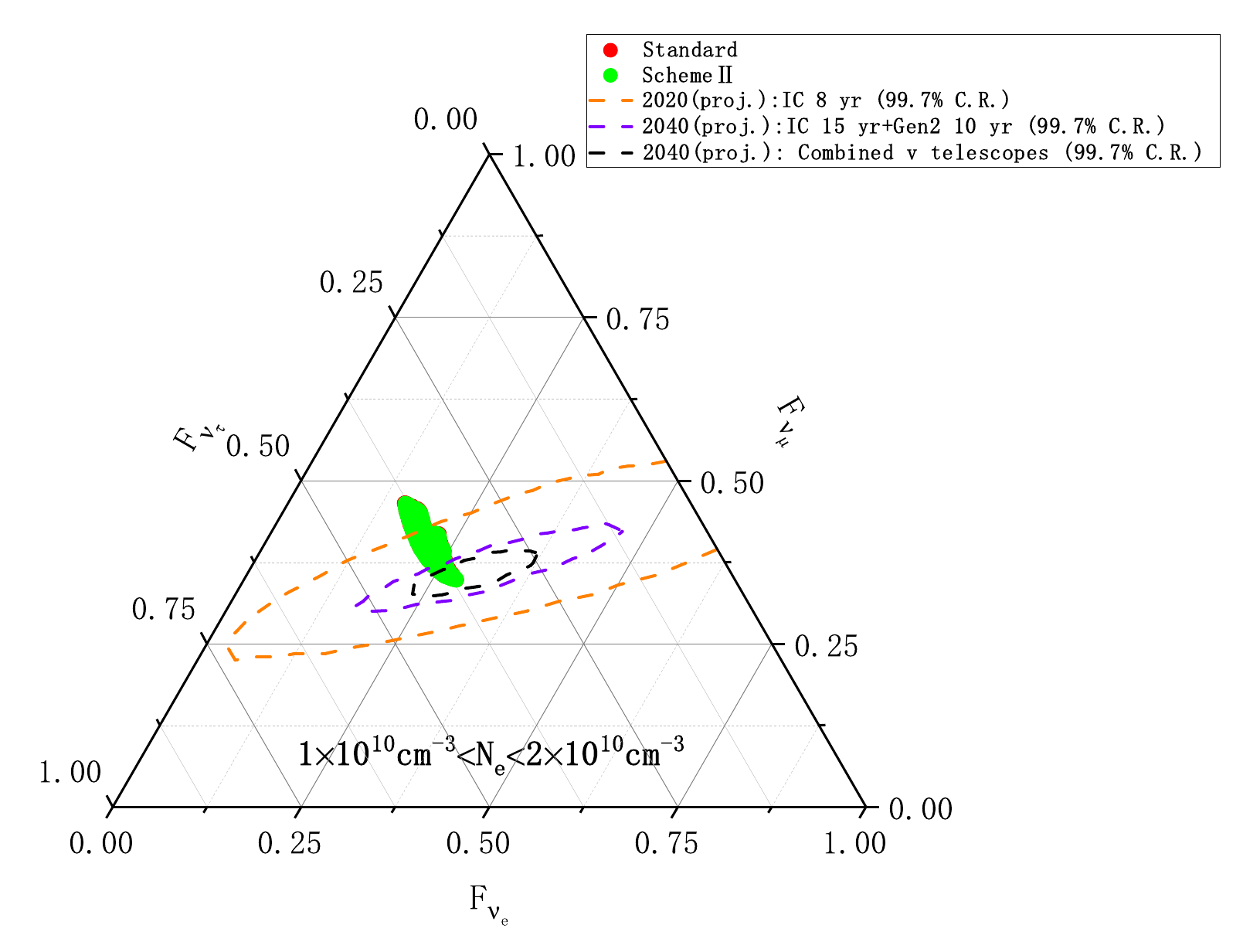}
\includegraphics[width=0.45\textwidth]{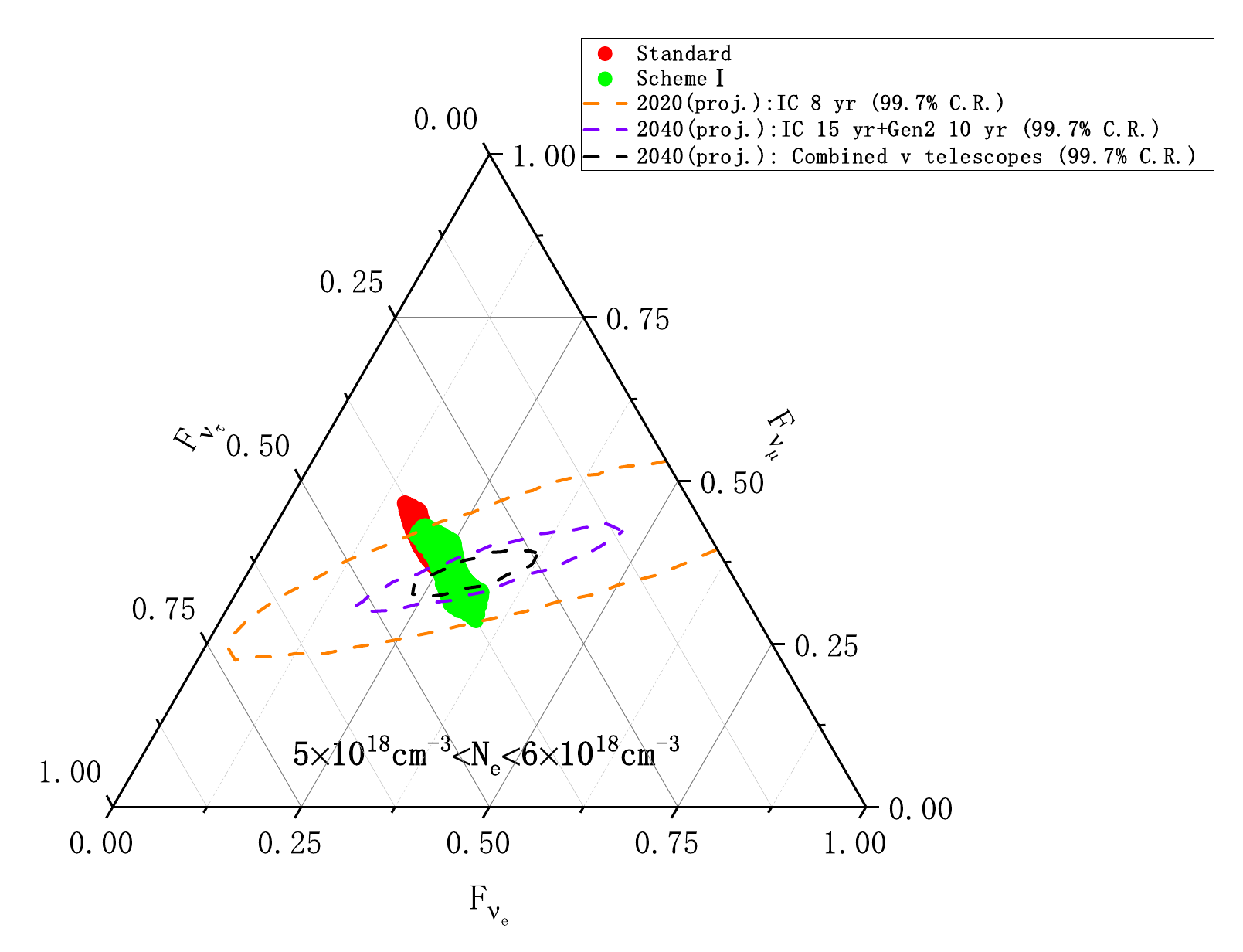}
	\includegraphics[width=0.45\textwidth]{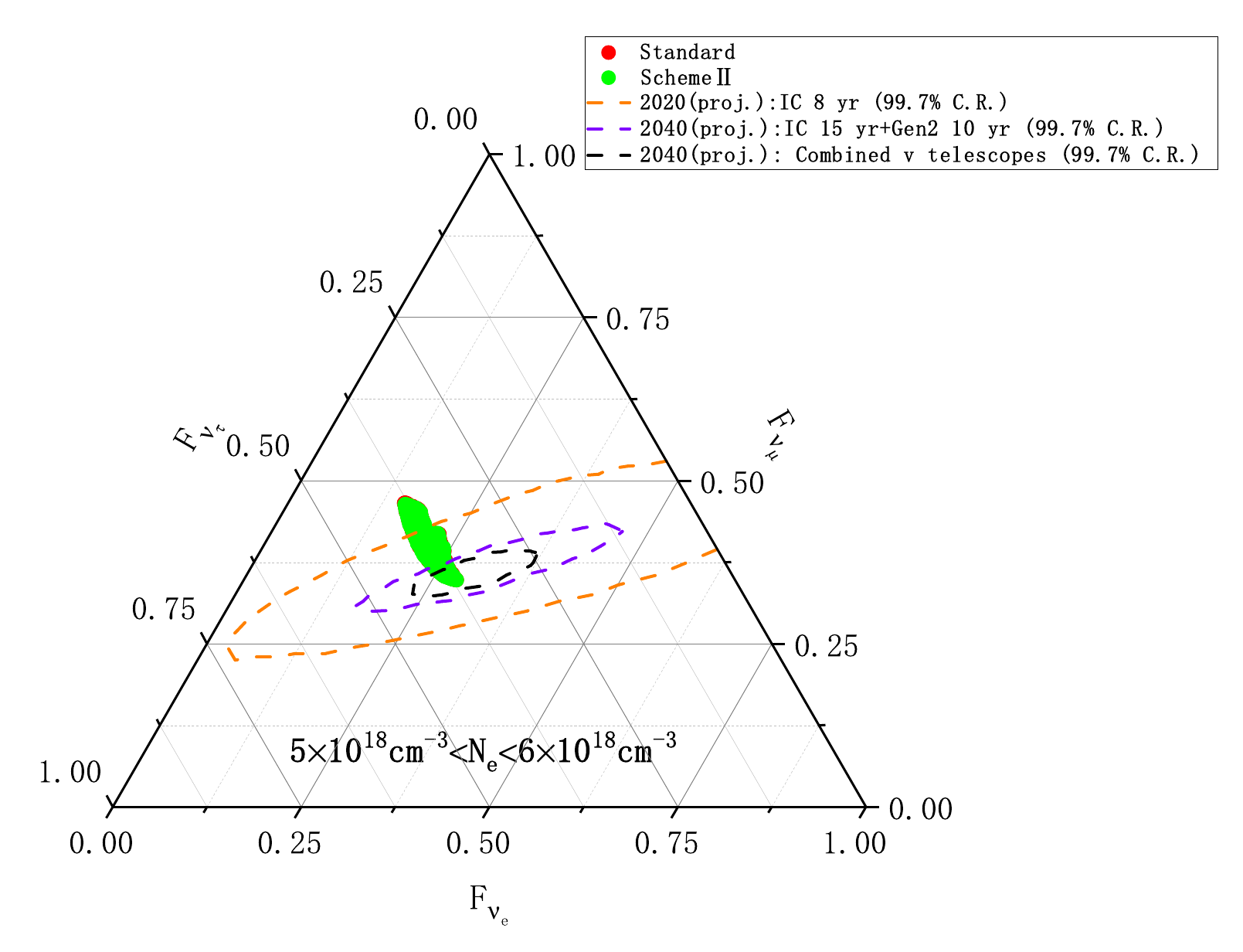}
	\caption{\label{fig:2} Ternary plot of the flavor ratio of HANs at Earth in the case of $\mu^{\pm}$ damping source. The neutrino energy is in the range [100TeV, ~1PeV], Neutrino oscillation parameters are taken in the $3\sigma$ allowed range of NuFIT 6.0 for the global fit data(NO) \cite{Esteban:2024eli}.  The orange dashed lines encompasses the 2020 $3\sigma$ C.R. with the $\pi$ decaying source based on IceCube\cite{21}.  The purple dashed line covers the 2040 $3\sigma$ C.R. based on IceCube and IceCube-Gen2\cite{21}. The black dash lines  denotes the $3\sigma$ C.R. boundary based on the TeV-PeV neutrino telescopes available in 2040\cite{15}. }	
\end{figure}
\begin{figure}\label{fig:3}
	\centering
	\includegraphics[width=0.45\textwidth]{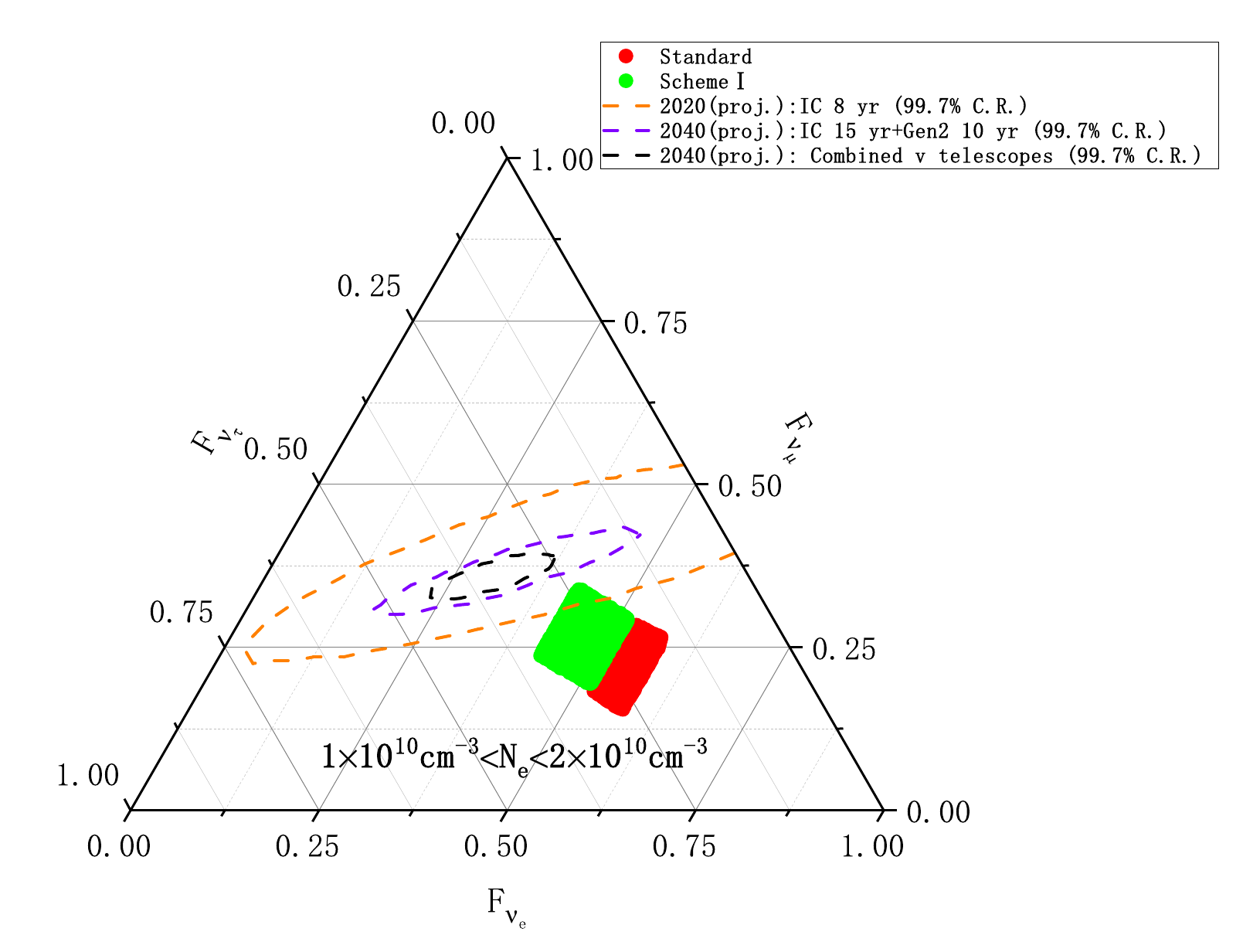}
	\includegraphics[width=0.45\textwidth]{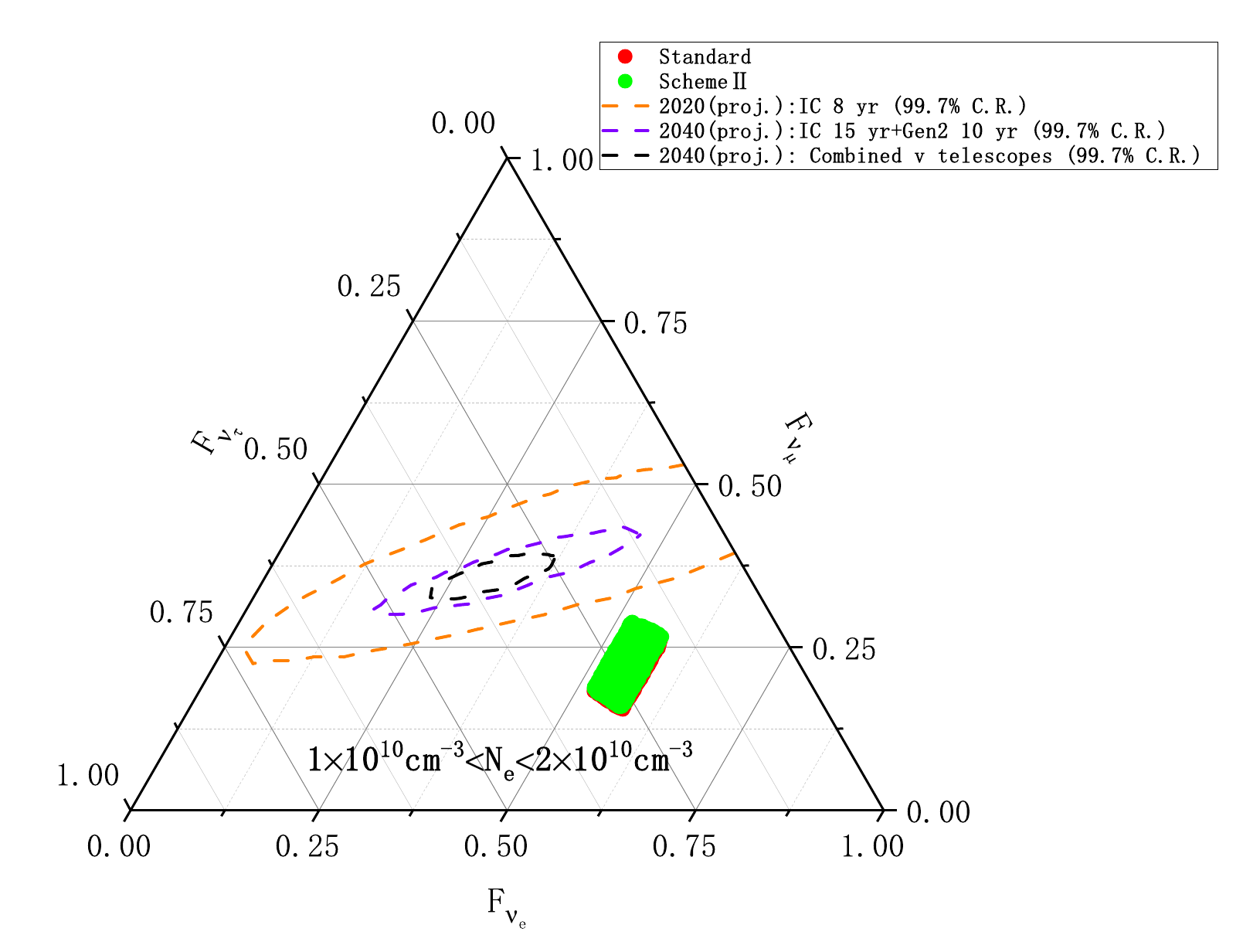}
\includegraphics[width=0.45\textwidth]{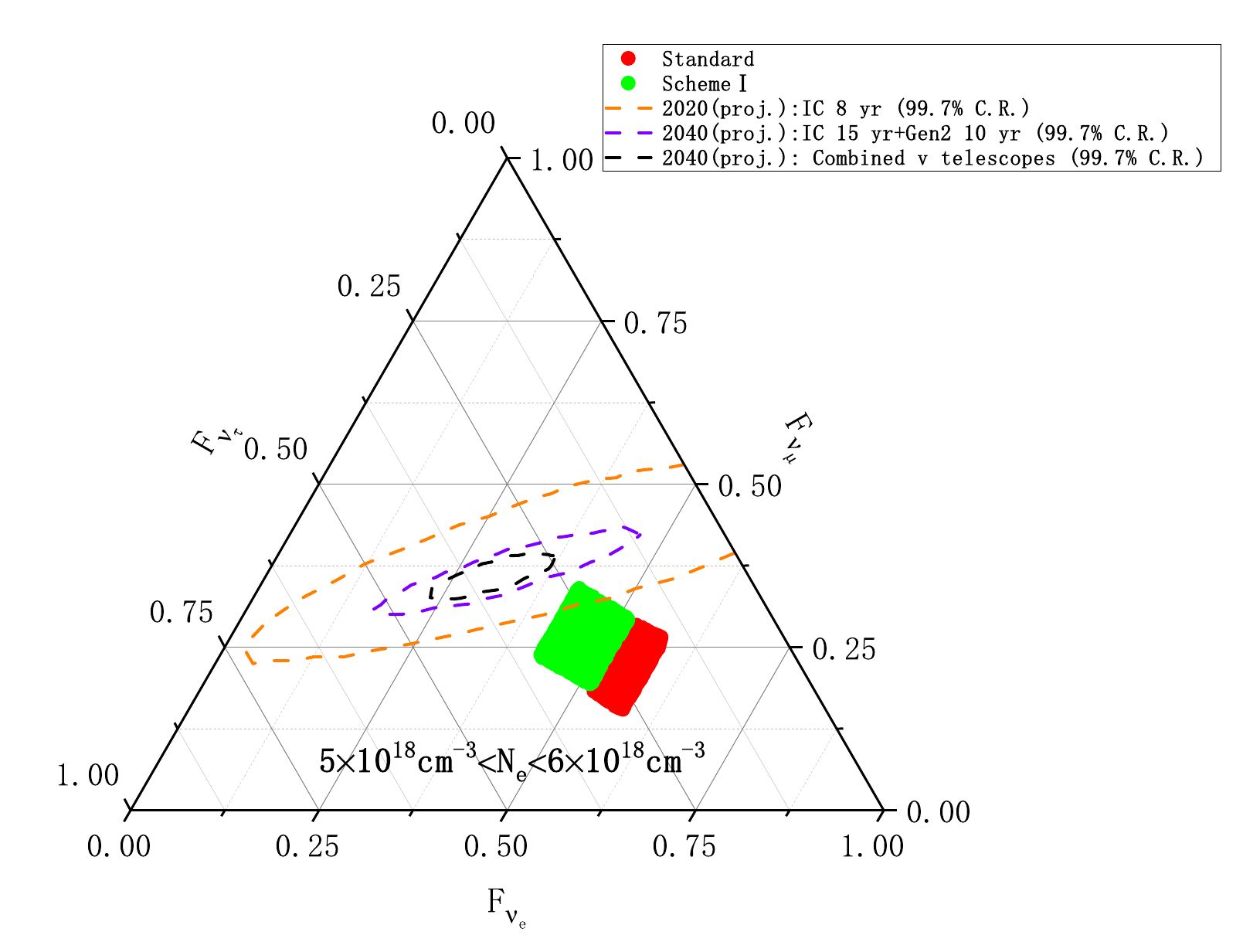}
	\includegraphics[width=0.45\textwidth]{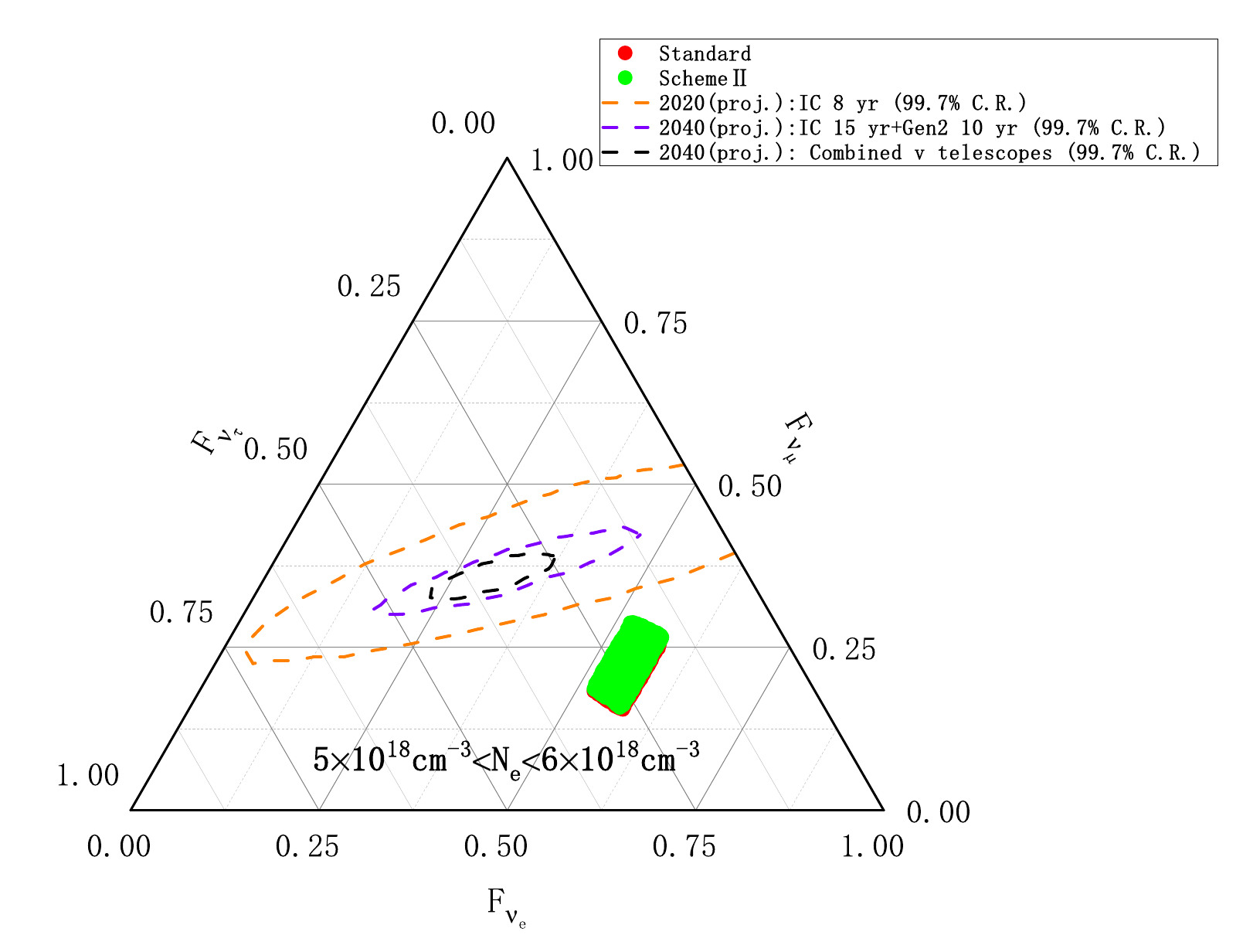}
	\caption{\label{fig:3} Ternary plot of the flavor ratio of HANs at Earth in the case of neutron decaying source. The conventions of the neutrino parameters are the same as those in Fig.\ref{fig:2}. }	
\end{figure}
\begin{figure}\label{fig:4}
	\centering
	\includegraphics[width=0.45\textwidth]{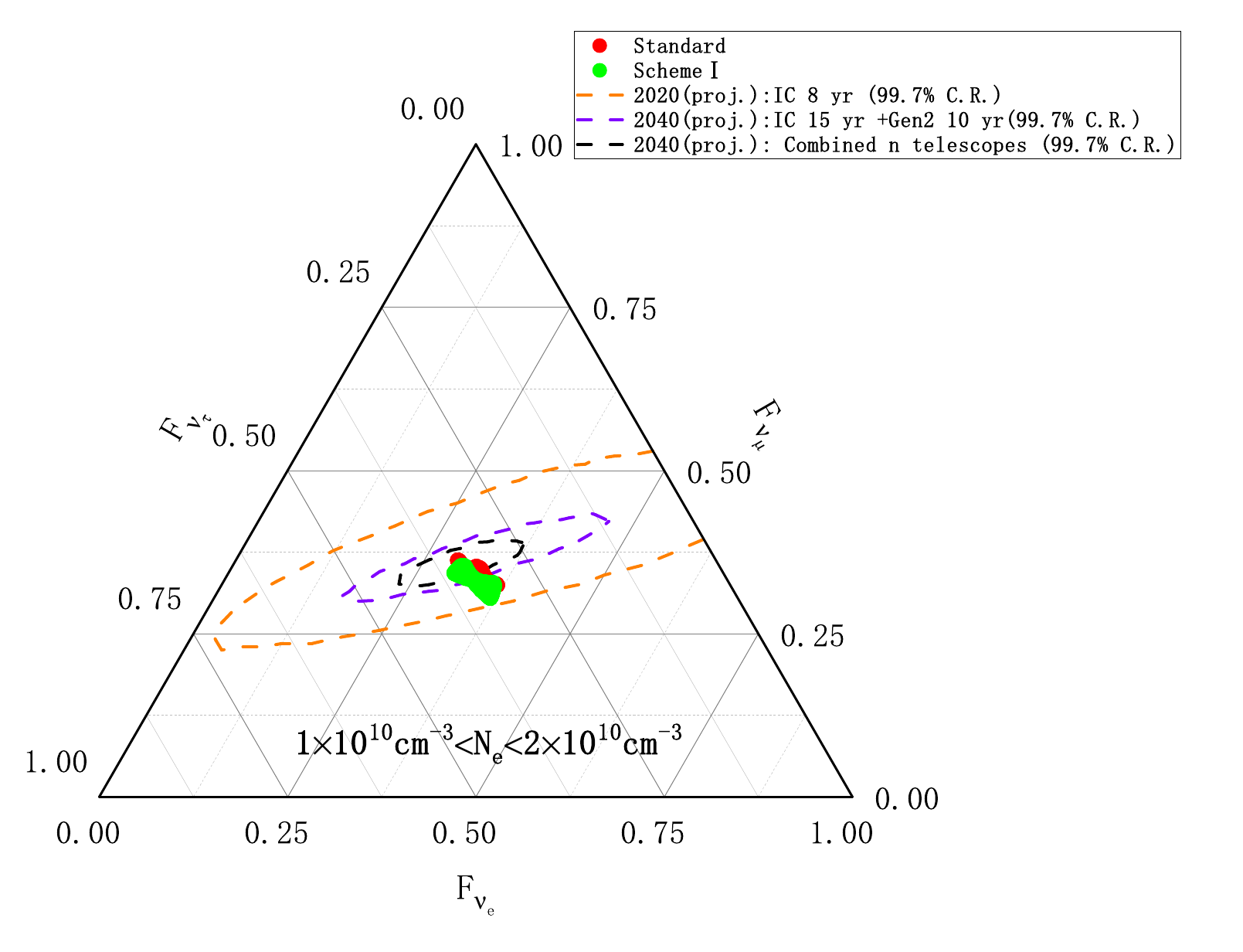}
	\includegraphics[width=0.45\textwidth]{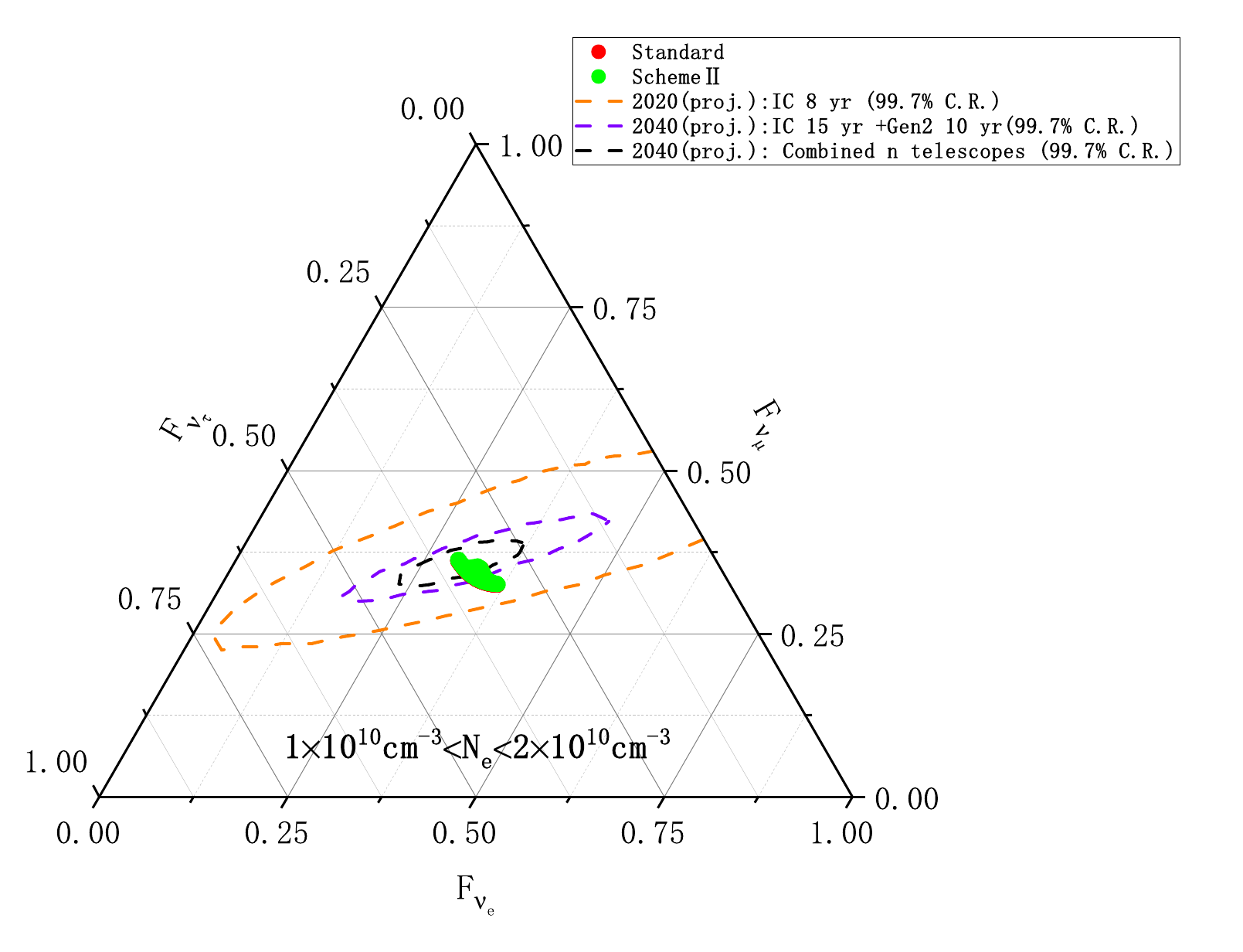}
\includegraphics[width=0.45\textwidth]{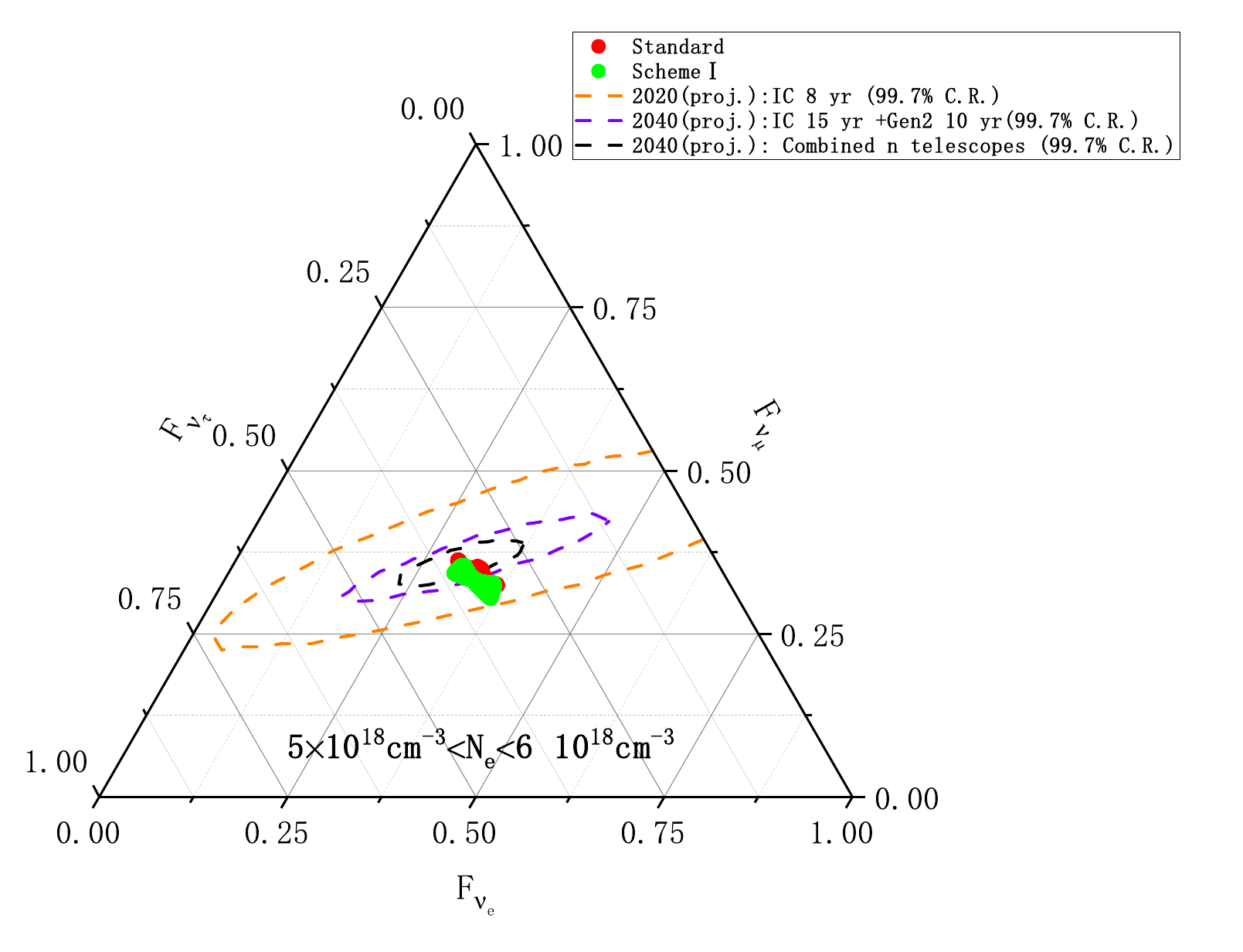}
	\includegraphics[width=0.45\textwidth]{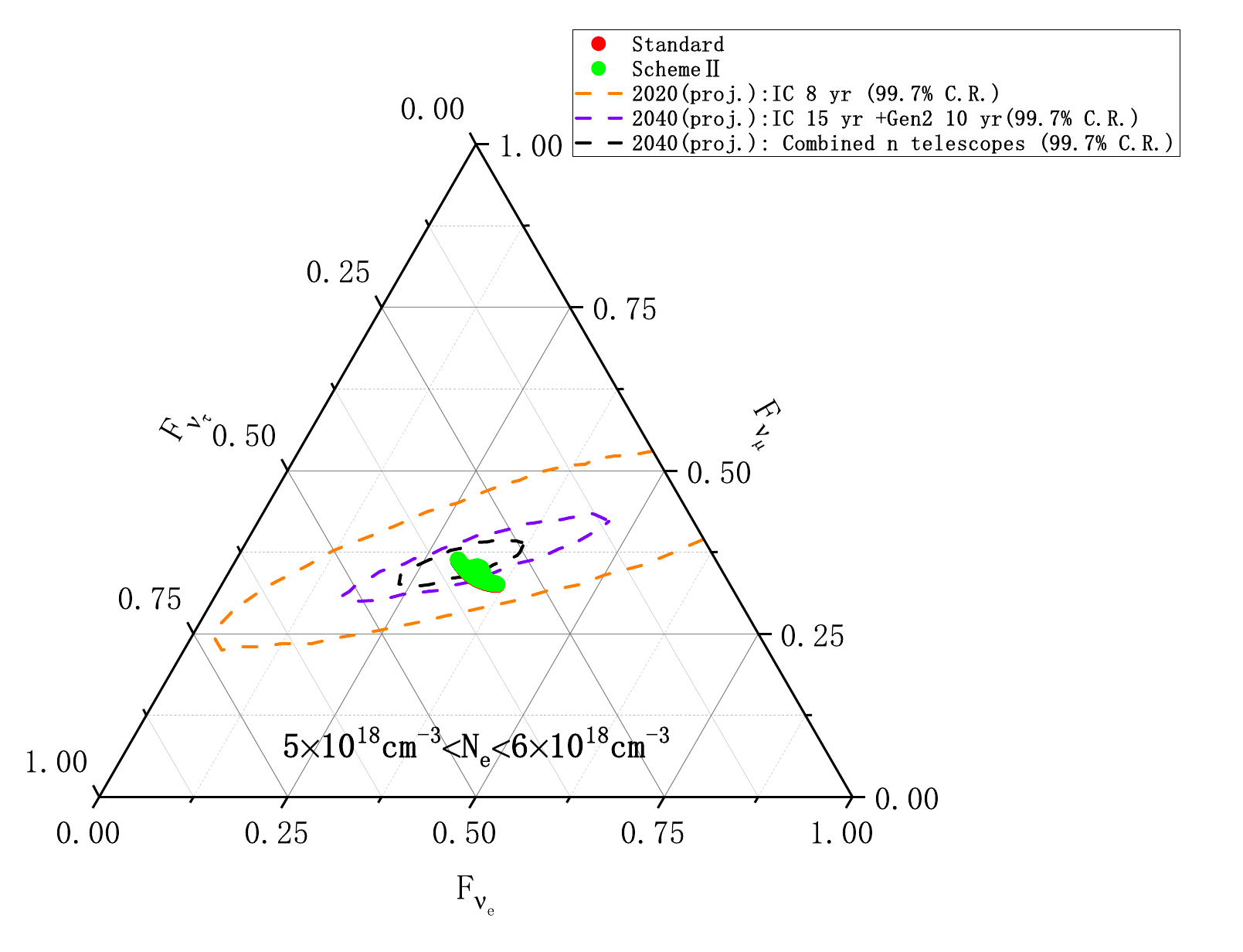}
	\caption{\label{fig:4} Ternary plot of the flavor ratio of HANs at Earth in the case of $\pi^{\pm}$ decaying source. The conventions of the neutrino parameters are the same as those in Fig.\ref{fig:2}. }	
\end{figure}

The plots manifest that the matter effect on the flavor ratio is dependent on the propagation schemes and the neutrino sources.
For the Scheme \uppercase\expandafter{\romannumeral2}, the impact of matter on the ratio
is negligible in the cases of the three typical sources.
For the Scheme \uppercase\expandafter{\romannumeral1}, the matter influence is obvious in the cases of muon damping and neutron decaying sources. In the case of $\pi^{\pm}$ decaying source,
although the matter effect is moderate, it is difficult to identify in the near future. For the both schemes, the predictions of the flavor ratio at low densities cannot be discriminated from
the results at high densities.
Furthermore, we find that other propagation schemes based on modifications of our schemes can bring similar observations in the ternary plots.

\section{Conclusions}

A precise measurement of the flavor ratio of HANs could be available in the future.
Testing the matter effects on the flavor transition of HANs by the next-generation neutrino telescopes deserves considerations.
The matter effects are dependent on the propagation schemes of the HANs and the production sources.
We considered two typical propagation schemes of HANs, namely propagation in the medium with a constant electron density and propagation in the medium with an adiabatically varying electron density.
On the basis of the recent updated data of HESEs from IceCube, we performed a simplified likelihood analysis on the electron density $N_{e}$ in the two schemes.
We find that the constraints from  HESEs on $N_{e}$ are weak in the both schemes. Thus, a large range of $N_{e}$  can be taken into account for the environment of HANs.
Based on the proposed electron densities, the matter impacts on the flavor ratio at Earth were examined with the three typical sources.
For the adiabatic Scheme \uppercase\expandafter{\romannumeral1}, the identification of the matter effects in the cases of muon damping and neutron decaying sources
is promising at at low and high electron densities.
For the constant density Scheme \uppercase\expandafter{\romannumeral2}, the matter effects are negligible, irrespective of the sources and the values of $N_{e}$.

\vspace{0.08cm}

\acknowledgments
We thank Qiu-Xia Yi for the help in the revision of the ternary plots. This work is supported by the National Natural Science Foundation of China under grant No. 12065007.

\bibliography{refs}

\end{document}